\documentclass{iau_FM} \usepackage{graphicx}
\title[FM 5.~~Blind Search for Variability] 
{Blind Search for Variability in Planck Data}

\author[J.P. Rachen, E. Keihanen \& M. Reinecke] 
{J\"org P. Rachen$^1$, Elina Keih\"anen$^2$ \and Martin Reinecke$^3$}

\affiliation{ 

$^1$ Department of Astrophysics/IMAPP, Radboud University Nijmegen, The Netherlands\\[\affilskip] 

$^2$ Department of Physics, University of Helsinki, Finland\\[\affilskip] 

$^3$ MPI for Astrophysics, 85741 Garching, Germany}

\pubyear{2015}
\jname{Astronomy in Focus, Volume 1} 
\editors{Piero Benvenuti, ed.}
\begin{document}

\maketitle

\begin{abstract} The sky is full of variable and transient sources on all time
scales, from milliseconds to decades. \textit{Planck}\/'s regular scanning strategy makes
it an ideal instrument to search for variable sky signals in the millimetre and
submillimetre regime, on time scales from hours to several years. A precondition
is that instrumental noise and systematic effects, caused in particular by
non-symmetric beam shapes, are properly removed. We present a method to perform
a full sky blind search for variable and transient objects at all Planck
frequencies.  
\end{abstract}

\firstsection 

\section{Introduction}

\textit{Planck} (\cite[Tauber et al. 2010]{Tauber2010}, \cite[Planck
Collaboration I. 2011]{PlanckI2011}) provides 8 full sky surveys at the
frequencies of its LFI instrument (30, 44, and 70 GHz), and 5 full surveys at
the 6 HFI frequencies in the range 100--857 GHz. The satellite rotates with 
1\,rpm around an axis kept fixed for a pointing period (PID) of about 50
minutes, then shifts along the ecliptic by 2'. This keeps a point source near
the ecliptic in the main beam (2 FWHM) for about $5{-}30$ PIDs (depending on
frequency) for each survey; for sources near the ecliptic poles the coverage can
be much larger. Analysing the time information in Planck data for a particular
sky direction thus allows to search for variability in the sky signal.

\vspace{-9pt}

\section{Mapping of time ordered information}

\begin{figure}[t] 
\begin{center}
\includegraphics[width=0.8\textwidth]{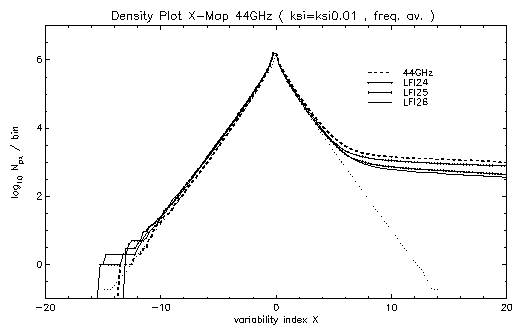} 
\caption{Histogram of an X-map for the LFI 44\,GHz detectors for $\xi=0.01$,
and detector noise $n_{\sf s}$ as given in \cite[Planck Collaboration II.
(2014)]{PlanckII2014}, thin dotted lines show the distribution expected for pure
Gaussian noise. The tail to large values of $X$ indicates the presence of true
sky variability. A tail to large negative values of $X$ would indicate an
overestimation of $\xi$, an offset of the peak from $X=0$ an incorrect
estimation of $n_{\sf s}$.} 
\label{fig1} 
\end{center} 
\end{figure}

Variability mapping is based on four-dimenional Healpix (\cite[G\'orski
et\,al.~2005]{Healpix}) constructs called 4D-maps, which record for every sky
pixel $k$ all contributions of a given detector at times $t_j$ and beam
orientation $b_j$, where the index $j$ refers to Planck PIDs. To
construct an average sky signal $S_k$ free from beam orientation effects, we use
the \textit{ArtDeco} beam deconvolution code (\cite[Keih\"anen \& Reinecke
2012]{ArtDeco}). A variability map is then a two-dimensional Healpix map of a quantity 
$$
X_k = \frac12\,\sf{sgn}(\hat\chi^2_k{-}\,1)\,N_k\,\Big[\hat\chi^2_k - 1 - \ln
\hat\chi^2_k\Big] 
$$ 
where\\[-5pt] 
$$ \hat\chi^2_k \equiv \frac{\chi^2_k}{N_k} = {1\over
N_k} \,\sum_j\, {(I_{kj} - (S{*}b)_{kj})^2 \over \sigma_k^2} 
\quad\qquad{\rm with}\qquad 
\sigma_{kj}^2 = \sigma_{\sf n}^2 + \xi^2\,(S{*}b)_{kj}^2\;. 
$$ 
$N_k$ is the number of entries for pixel $k$ in the 4D signal map $I_{kj}$,
$(S{*}b)_{kj}$ is the beam-reconvolved average 4D-map based on the \textit{ArtDeco} map
$S_k$, $\sigma_{\sf n}$ is the detector white noise, and $\xi$ combines all
instrumental fluctuations which factor on the signal (e.g., calibration,
inaccuracies of the beam model). The definition of $X$ is motivated by the
Chernoff bound on the CDF $P_{\!N}(\chi^2)$, with \mbox{$X \le - \ln (1-P_{\!N}(\chi^2))$}
for $\hat\chi^2 \ge 1$, and $X > \ln P_{\!N}(\chi^2)$ for $\hat\chi^2 < 1$. The
distribution of X-values over a sky map is indicateive not only of variability,
but also to incorrect estimations of noise and instrumental variations (see
Fig.~\ref{fig1}).

\vspace{-9pt}

\section{Time resolved Planck fluxes, status and outlook}

As the analysis of time variations in the sky is essentially background-free,
our method provides a way to extract time resolved Planck fluxes down to a time
resolution of a few hours. If the position of a variable point source is known,
the residual 4D-map, $R_{kj} = I_{kj} - (S{*}b)_{kj}$, can be beam-deconvolved
to the source position by a simple division and thus provide an estimate of the
flux variation, $\langle \Delta S(t_j)\rangle_k = R_{kj}/b_{kj}$, where $b_{kj}$
is a \textit{beamfactor map} expressing the measured flux of a unit emitter at
the source position in a beam centered at pixel $k$ and time $t_j$. This method
is used, e.g., to extract time-resolved Planck fluxes for co-eval monitoring of blazars
with the F-GAMMA program (\cite[Fuhrmann et al. 2007]{F-GAMMA},
\cite[Rachen et al. 2015]{BlazarPoster}).

Our analysis tools work on unified data structures, which ensures that all
methods for variability analysis and flux extraction are applicable to LFI and
HFI data in the same way. Interfaces for intitial 4D-mapping are in place at
both the LFI and HFI DPC. We expect that all Planck frequencies will be analysed
for variability, and science results prepared for publication well before the
Planck Legacy release.

\end{document}